# Double Clad Antiresonant Hollow Core Fiber and Its Comparison with other Fibres for Multiphoton Micro-Endoscopy

Marzanna Szwaj[1,2,3], Ian A Davidson[1], Peter B Johnson[2,3], Greg Jasion[1], Yongmin Jung[1], Seyed Reza Sandoghchi[1], Krzysztof P Herdzik[1,2,3], Konstantinos N Bourdakos[2,3], Natalie V Wheeler[1], Hans Christian Mulvad[1], David J Richardson[1], Francesco Poletti[1]*, Sumeet Mahajan[2,3]*

[1]*Optoelectronics Research Centre, University of Southampton, SO17 1BJ, UK*
[2]*Institute for Life Sciences, University of Southampton, SO17 1BJ, UK*
[3]*School of Chemistry, University of Southampton, SO17 1BJ, UK*
*frap@orc.soton.ac.uk
*S.Mahajan@soton.ac.uk

Label-free and multiphoton micro-endoscopy can transform clinical histopathology by providing an in-situ tool for diagnostic imaging and surgical treatment in diseases such as cancer. Key to a multiphoton-imaging based micro-endoscopic device is the optical fiber, for the distortion-free and efficient delivery of ultra-short laser pulses to the sample and effective signal collection. In this work, we study a new hollow-core (air-filled) double-clad anti-resonant fiber (DC-ARF) as a potent candidate for multiphoton micro-endoscopy. We compare the fiber characteristics with a single-clad anti-resonant fiber (SC-ARF) and a solid core fiber (SCF). While the DC-ARF and the SC-ARF enable low-loss (<0.2 dBm$^{-1}$), close to dispersion-free excitation pulse delivery (<10% pulse width increase at 900 nm per 1 m fiber) without any induced non-linearities, the SCF resulted in spectral broadening and pulse-stretching (> 2000% of pulse width increase at 900 nm per 1 m fiber). An ideal optical fiber endoscope needs to be several meters long and should enable both excitation and collection through the fiber. Therefore, we performed multiphoton imaging on endoscopy-compatible 1 m and 3 m lengths of fiber in the back-scattered geometry, wherein the signals were collected either directly (non-descanned detection) or through the fiber (descanned detection). Second harmonic images were collected from barium titanate crystals as well as from biological samples (rat tail tendon). In non-descanned detection conditions, the ARFs outperformed the SCF by up to 10 times in terms of signal-to-noise ratio of images. Significantly, only the DC-ARF, due to its high numerical aperture (0.45) and wide-collection bandwidth (>1 μm), could provide images in the de-scanned detection configuration desirable for endoscopy. Thus, our systematic characterization and comparison of different optical fibres under different image collection configurations, confirms and establishes the utility of DC-ARFs for high-performing label-free multiphoton imaging based micro-endoscopy.

## Introduction

Successful treatment of cancer can benefit immensely from early detection and accurate diagnosis of pathological abnormalities. Currently, the definitive diagnosis of cancer tumors still relies on histopathology as the "gold standard" for evaluating the morphological and molecular changes in tissue sections [1]. In this process, tissue samples are obtained by biopsy procedures, and then sectioned and stained to undergo microscopic examination by a histopathologist. Sampling errors due to limited real-time information and inability to compare diseased areas with their healthy counterparts *in situ* can therefore limit the diagnostic accuracy of histopathological analysis. Moreover, it is time-consuming and labor-intensive. Additionally, for intra-operative surgical guidance, especially for tumor margin detection, a rapid, real-time, non- or at least minimally-invasive micro-endoscopic imaging method is needed.

Label-free multi-photon microscopy (MPM) is a promising set of techniques for real-time imaging of biological samples in their native state. MPM often uses near infrared wavelengths that cause less damage to tissues and have a deeper penetration range than their visible counterparts [2, 3]. The key challenge to take advantage of MPM is to transit from bulky, laboratory instrumentation into a miniaturized and portable endoscopy system that is compact and flexible enough to find application in the clinical environment. Such a platform could offer slide-free, label-free, real-time, *in vivo* imaging of cancer tumors and open the path to the development of *in situ* "digital histopathology".

Portability and flexibility in such an endoscopic system is accomplished by utilization of optical fibers. However, the signal intensity of MPM is inversely proportional to the pulse-width of the laser [4], therefore ultra-short (pico-/femto-second) laser pulses are applied. Hence, fiber selection is critical to deliver ultra-short pulses unaffected by pulse broadening due to chromatic dispersion and non-linearity respectively. Chromatic dispersion causes pulse chirp (i.e. the wavelength to vary with time across the broadened pulse) while non-linear effects lead to alteration of the phase and the generation of light at new wavelengths (which can be a problem in its own right but also lead to increased temporal broadening) through self-phase modulation (SPM) or four wave mixing (FWM). In addition to minimizing these effects, the chosen fiber waveguide should have a low attenuation and wide transmission window to realize certain MPM modalities such as second harmonic generation (SHG) and coherent anti-Stokes Raman scattering (CARS)[5].

The main challenge of ultra-short pulse delivery via a solid core fiber (SCF) is temporal pulse broadening, as was demonstrated in an MPM system with two-photon fluorescence [6]. The primary reason for this is the large group velocity dispersion (GVD) and power dependent non-linearity related to propagation of pulses in a silica fiber core with its small core diameter, respectively. Several solutions have been proposed such as the use of shorter fiber lengths [7], or use of large mode area fiber [8, 9] so that pulse alteration is negligible. Polarization maintaining fiber (PMF) with dual wavelength wave-plate (DWW) [10, 11], and a microfabricated long-pass optical filter [12] have been shown to reduce the non-linear four-wave mixing (FWM) noise in MPM techniques such as CARS that require multiple laser wavelengths for excitation. The aforementioned PMF-based schemes maintain the state of polarization (SOP) of each of the laser beams and couple them orthogonally along the slow and fast axes of the fiber, and at the distal end of PMF, DWW realigns polarization to their original, pre-fiber condition. Dispersion compensation schemes based on a pair of gratings to pre-chirp the pulses have also been proposed [13–16]. However, all these methods raise the complexity of the system and add bulk to the micro-endoscope design which ideally needs to be handheld and/or miniature.

Hollow core fiber (HCF) technology brought hope of more efficient and background-free ultra-short, high power delivery of laser pulses. In an HCF non-linear effects are suppressed by several orders of magnitude due to propagation in an air core and as such are negligible. Hollow core photonic band gap (HC-PBG) fibers were tested in fiber-coupled systems and shown to be a good candidate for non-linear endoscopy [17 – 21]. However, their narrow transmission bandwidth (≤100 nm) [22] can be a limitation for MPM modalities where the excitation and signal wavelengths can span >400 nm. Additionally, GVD changes rapidly with wavelength, with dispersion changing from the anomalous to normal regime across the transmission band [23]. Thus, a specially-designed chirped photonic crystal fiber (CPCF) was successfully applied for MPM without any significant wavelength-dependent distortion [24] although the transmission bandwidth was still limited. Further advances have led to the development of "Kagome" – lattice hollow core photonic crystal fiber (HC-PCF). In Kagome HCF the cladding microstructure does not create an optical bandgap and propagation occurs through two combined guidance mechanisms, namely the inhibited coupling (IC) of core and cladding modes, and the effect of an anti-resonant reflection optical waveguide (ARROW) mechanism from the glass core boundary [25]. Kagome fibers are typically characterized by a much larger core diameter (<20 μm) [26] compared to an HC-PBG (~5.6 μm) [22] and have a broader bandwidth of >150 nm [27] compared to ≤100 nm in a HC-PBG [22] but at the expense of increased attenuation. Dispersion in this fiber has a shallow slope with a value of -0.5 $ps^2$ $km^{-1}$ at 800 nm for a core diameter of 30 μm [28]. Nevertheless, a Kagome fiber with an attenuation level of ~3 dB/m at 800 nm was successfully applied in multiphoton micro-endoscopy with high power fs laser excitation pulses [26].

Emergence of the hypocycloid-core Kagome HC PCFs [27] marked the beginning of a new era of tubular air-filled core fibers with simplified geometry while greatly reducing fabrication complexity and therefore, distinguishing them from other microstructured HCFs. Their design is based on an air core surrounded by capillaries. The ARROW based guidance mechanism has mostly been accepted by the fiber optic community, and hence tubular fibers frequently are named anti-resonance fibers (ARFs) [29-30]. The glass membrane thickness of the capillaries surrounding the core determines the spectral location of an optical resonance. Recently, a low transmission loss of $1.45 \pm 0.15$ dB $km^{-1}$ at 850 nm was reported [31] and advances in HCF technology are ongoing rapidly. ARFs offer a broad transmission bandwidth which can span more than an octave [32], a spectrally smooth low loss curve and have a large core diameter size (>~25 μm). Additionally, the higher damage thresholds offered by ARFs allow high power pico- and nanosecond pulse delivery at different wavelengths [33-34]. One such ARF has been successfully used to couple fs pulses into a multiphoton microscope to obtain SHG and TPF images of *ex vivo* tissue [35]. Nevertheless, typical ARFs are characterized by a low numerical aperture (<0.05), while for non-linear imaging, especially with techniques such as CARS, a large (~0.5 or higher) NA is highly desirable.

In addition to delivering ultra-short pulses to the sample without spectral or temporal distortion, the fiber for endoscopy must be able to collect the back-scattered MPM signal. Therefore, there is a growing interest in adaptation of the fiber cladding to obtain a higher NA in the design of fibers for micro-endoscopic systems. A single-fiber-based endoscope with capacity for simultaneous pulse delivery and backward signal collection has been investigated and explored by many researchers. Commercially available [36-38], or specially customized double-clad solid core fibers (DCFs) [39] were tested for remote non-linear microscopic application. Their common feature is a large inner-cladding and a single mode core that provided the best imaging resolution. The same principle has been applied with a double-clad photonic crystal fiber (DC-PCF) fabricated with 3.5 μm and 188 μm diameter of inner core and outer cladding, respectively [40]. However, in the above, the challenge of distortion-free pulse laser delivery without a GVD pre-compensation scheme remains. HCF technology provides a solution with negligible GVD and non-linearity, but an intrinsically large core diameter adversely affects the image resolution. This can be seen in the case of a double-clad HC-PBG fiber applied to a coherent Raman "endoscope-like" scheme [21]. An alternative approach was presented for a DC-Kagome [26] and a DC-ARF [41] in which a silica micro-bead was attached to the hollow core to increase NA and resolution. Most recently, Kudlinski et al. [41] described a new DC-ARF customized with an extra-large clad area and layers of low and high refractive index jacket. The reported DC-ARF has a wide and low-loss bandwidth from 700 to 1500 nm, which is ideal for non-linear micro-endoscopy. The DC-ARF uses the typical tubular architecture with 8 inner capillaries surrounding a fibre core of 30 μm and the average capillary thickness is about 250 nm. The distance between capillaries is 5±1 *μ*m and their diameter is 12.6±0.7 *μ*m. While TPF and SHG images have been shown, the details of the micro-endoscopic system, its configuration or imaging performance of different potential configurations were not presented.

Here we use an alternate design of a double-clad DC-ARF with non-touching tubes and demonstrate its performance with label-free multiphoton (SHG) imaging in two different backward signal collection configurations corresponding to non-descanned and descanned detection (NDD and DD). In the NDD configuration, only the excitation is through the fiber, while in the DD configuration both the excitation delivery and the signal collection is through the fiber. We present the characterization of our new DC-ARF which shows that it is suitable for high-power, ultra-short pulse transmission and backward collection of the signal as required in a multiphoton micro-endoscope. The large cladding area and high NA provide for increased signal collection efficiency. Furthermore, we compare the new DC-ARF with an SC-ARF of similar geometry and characteristics but without the double-cladding, and with a standard

SCF. The SC-ARF and SCF are only able to guide light in the core region. All three waveguides are evaluated based on their ability to deliver a distortion-free ultra-short laser pulse to the sample, to collect the back-scattered signal in SHG imaging. The effectiveness and performance for potential endoscopic application is demonstrated under both NDD and DD configurations for all three fibers using both 1 and 3 meter long fibers. Under our experimental conditions the DC-ARF was the only fiber that allowed imaging in the DD configuration, which is the most compatible with micro-endoscopy. Our work provides a systematic study comparing the performance of three different fibres under different collection configurations and confirms DC-ARF as a promising candidate for high quality micro-endoscopy with multi-photon imaging techniques.

## Hollow core double clad anti-resonant fibre (DC ARF) fabrication and loss characterization

The DC-ARF used in this work was fabricated in a stack and draw process with Heraeus F300 high purity fused silica glass tubes, and coated with a low refractive index ($n$ = 1.376 at 852 nm) polymer jacket, PC-373-AP, to obtain the contrast required for guidance in the glass inner cladding surrounding the capillaries. The low-index coating layer serves as an outer cladding for the back-collected light [32]. An optical image of the DC-ARF is shown in **Fig. 1a**. While the design principle of our DC-ARF fiber is similar to that of Kudlinski et al [41] our DC-ARF has an alternative construction, differing in the core diameter (26.3+/-0.3μm), the inner cladding (49.5 μm), the outer cladding (137 μm), number of tubes (7), tube spacing (4.5 μm) and membrane thickness (349±8 nm). With a total fiber diameter of 279 μm, our DC-ARF preserves a "standard" size of geometry, with only one layer of low refractive index coating increases fabrication replicability. Furthermore, by employing a fiber with smaller dimensions but a high cladding NA, better miniaturization of the system without compromising on the level of signal collection is achievable. The calculated total collection surface area is 12700 μm$^2$. At the low loss transmission wavelengths, the signal is confined to the air-core area by anti-resonance effects from the surrounding silica capillaries. However, light can also be simultaneously coupled into and propagated in the glass cladding by total internal reflection (TIR) **(Fig. 1b)**. Modelling showed that choosing seven tubes in the cladding was a good compromise between low attenuation, low bend-loss and large extinction of higher modes [42]. The NA in the core was experimentally measured to be 0.029 by launching light into the DC-ARF and measuring the output beam diameter at 1/e$^2$ of the maximum intensity at a known distance using a scanning-slit optical beam profiler (B2 290 -VIS, Thorlabs). Such an NA is mostly determined by the fundamental LP01 mode of the fiber, with the LP11 mode already strongly attenuated after short distances and higher order modes not present, as confirmed by S$^2$ measurements described previously [32]. This result is in agreement with other measured NA values of tubular fibers such as 0.036 for a 22 μm core diameter fibre [36] and 0.038 for a 15 μm core diameter fiber [35]. The NA of the cladding was estimated to be 0.45 based on the value of refractive index difference between glass and the coating material.

Attenuation was obtained by a cut-back measurement for 10 m and 40 m long fibers. A launch fiber (SMF-28) was coupled on one end to a white light source (WLS) and at the other end to the DC-ARF, maintaining the coupling conditions into the core. Attenuation spectra were recorded using an optical spectrum analyzer (OSA; Ando AQ – 6315A) with a spectral range of 400 – 1750 nm. The fundamental anti-resonant window of the DC-ARF is extremely broad **(Fig. 1c)** extending from around 750 nm to beyond 1750 nm. Furthermore, the DC-ARF can transmit visible light at wavelengths between ~425 nm to 675 nm in its core in its second anti-resonant window. The loss < 0.2 dBm$^{-1}$ is sustained over the 500 nm window between 800 nm – 1300 nm. The high loss region is for wavelengths which are in resonance with the tube membranes. At wavelengths where the resonance condition is satisfied, modes cannot stay confined in the hollow core, and they leak away through the glass membranes, creating spectral regions of high attenuation. In ARFs the spectral operational region can be modified by controlling the wall thickness [32].

For endoscopic applications, a relatively long fiber length of several meters is required, as it can bend during deployment. Hence, the optical performance of the DC-ARF was investigated for different bend radii. Applying the same launch fiber and measurement equipment as in the cut-back experiment, transmission spectra were obtained and analyzed for a 3 m long fiber. The results plotted in **Fig. 1 (d)** show that for bend radii larger than 10-15 cm and for wavelengths within the fundamental window the bend loss is barely detectable. At smaller bend radii and shorter wavelengths the bend loss starts becoming measurable. This bend-induced loss is similar to that studied by many groups [43-45], and is explained by the phase matching between the core mode and hollow tube modes.

## Suitability of fiber characteristics for multiphoton endoscopy: comparison of DC-ARF with SC-ARF and an SCF

Suitability of waveguides for multiphoton micro-endoscopy with non-linear imaging techniques can be evaluated based on three major criteria: broadband transmission, spectral pulse broadening due to non-linear effects, and dispersion due to laser propagation within the fiber. These characteristics determine whether the MPM excitation can be delivered effectively without loss and distortion. Additionally, a fiber for endoscopic applications should efficiently collect the often weak back-scattered signals. We investigated these characteristics for our DC-ARF and also compared its performance to an SC-ARF with a similar geometry (25 μm core diameter and seven tubes) but without the double clad, as well as with a SCF (single mode fiber, SMF-28, cut-off wavelength: 1260 nm). ARFs technology is known to offer strongly single mode performance after 10s of meters [32]. For the shorter lengths (≤ 3 m) of interest to this work, the fiber is still predominantly single mode, but with some expected small contribution from the LP11 mode which is not expected to affect negatively its performance. The single mode fiber (SMF) selection was based on its multimodal character below 1000 nm, that is, in the wavelength range of 800 – 1000 nm used in this work, the selected SMF operates as a multimode fiber and therefore matches the multimodal character of ARFs.

**Optical Transmission**

We measured the transmission in all three fibers using a white light source and an optical spectrum analyzer. We used a launch fiber to ensure single mode light confinement. A SMF-28 launch fiber was used to couple light into the HCF. Transmission spectra demonstrate that all three fibers have a good transmission across 800-1600 nm **(Fig. 2a)**. The SC-ARF and the DC-ARF, show similar transmission in the core, as expected as the microstructured regions are very similar. Between 600 to 750 nm the high loss indicates the on-resonance state of the light with the glass membranes. Furthermore, **Fig. 2 (a)** shows that the broadband guidance of DC-ARF cladding is very similar to that of the SCF (SMF-28) since both are essentially made of the same material. These results also support the concept of utilization of the DC-ARF cladding area for MPM signal collection in the visible spectral region.

**Non-linear spectral effects**

The presence of non-linear effects in the fiber manifests itself as broadening of the optical spectral width, which is proportional to the peak power. We tested this effect at two wavelengths: 810 nm and 900 nm and at two different laser powers input into the fiber: 20 mW and 60 mW. The wavelengths and powers were chosen based on the choice for MPM experiments with SHG imaging. We measured the spectra before and after each of the three fibers using a spectrometer (Ocean Optics Red Tide 650) to determine the effect due to non-linear processes with a 116 fs Ti:Sa pulsed laser. **Fig. 2 (b)** presents the results of spectral pulse broadening for selected wavelengths and power levels. The spectral full-width-half maximum (FWHM) bandwidth of the input fs pulses is ~10 nm. For the SC-ARF and DC-ARF, the output spectra are independent of power and comparable with the spectrum of the free-space fs laser. However in the SCF, broadening is clearly observed even at 20 mW. On increasing the power, self-phase modulation effects occur and cause splitting and broadening of the optical spectrum.

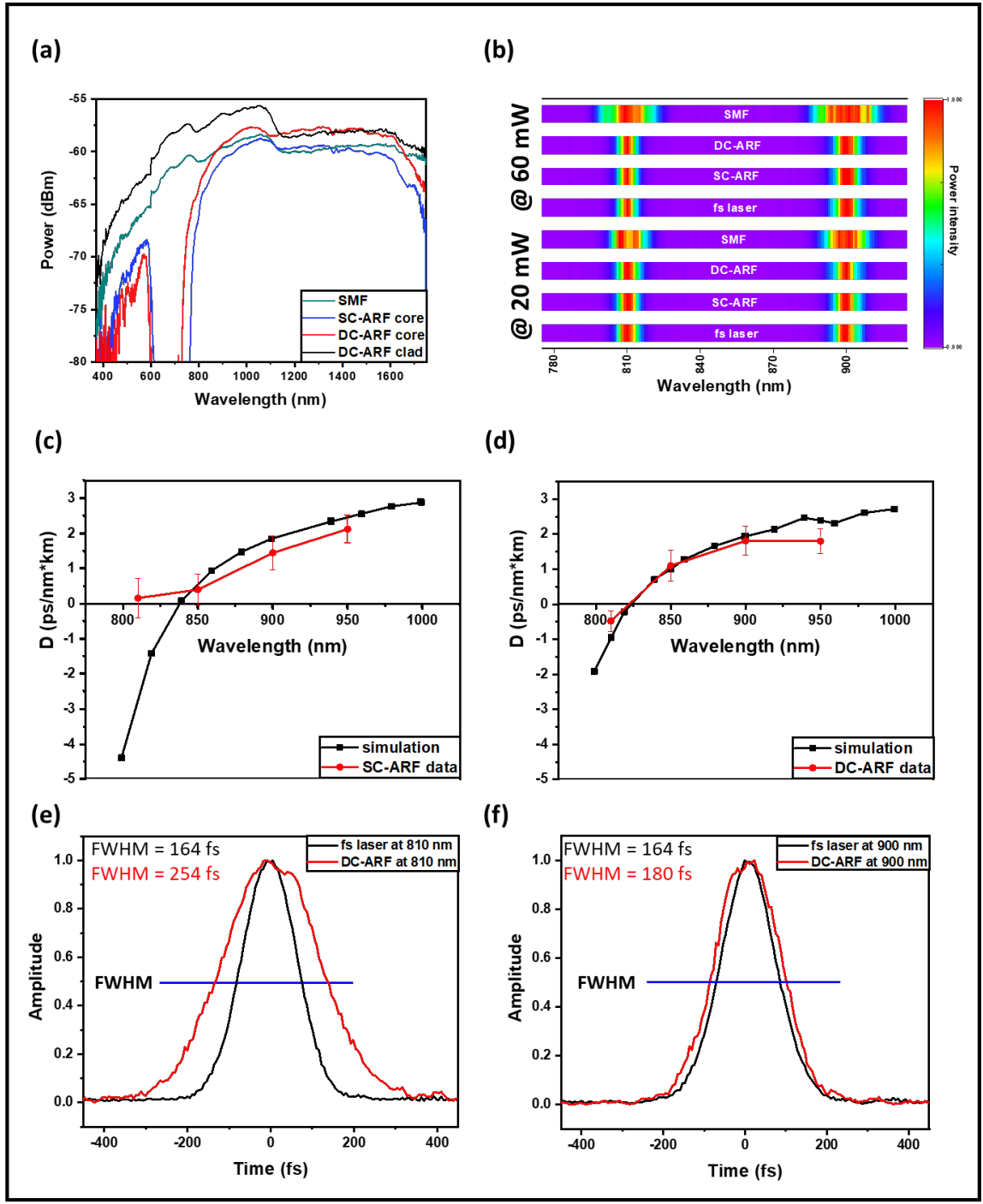

**Fig. 2 Characterization of fibers (a)** Measured transmission spectra in the core for SCF, SC-ARF, DC-ARF and also for the cladding in the DC-ARF **(b)** Spectra taken before and after delivery through 3 m long fibers of fs laser pulses at 810 nm and 910 nm, at a power of 20 mW (top four rows) and 60 mW (bottom four rows). Spectra were acquired with 10 ms integration time and normalized linearly **(c), (d)** Simulated GVD for SC-ARF and DC-ARF (black) vs measured results (red). **(e), (f)** Intensity autocorrelation traces of the transform-limited fs laser pulses (116 fs Ti:Sa) before and after 3 m of DC-ARF at 810 nm and 900 nm and 20 mW power. FWHM ($\Delta \tau_a$) of the auto-correlation are stated.

**Dispersion**

In order to evaluate the dispersion in the fiber, we measured the intensity autocorrelation trace with 20 mW average power transmitted through all three fibers at 810 nm, 850 nm, 900 nm, and 950 nm pulses using a Ti:Sapphire laser (116 fs, 80 MHz, MaiTai, Spectra physics). For both HCFs, the GVD parameter D is nearly zero in the 800 - 1000 nm region. However, fs laser pulses delivered in SCF experience a large anomalous dispersion. For the SC-ARF and DC-ARF, we conducted computational simulations based on the fibers' geometry and experimental attenuation data **(Fig. 2c, d)**. The experimental data are very close to the simulated results and any difference could be attributed to minor differences in actual fiber geometry. The D value of SC-ARF at 810 nm is much larger than expected and could be a result of operating near the edge of the transmission window for this wavelength as higher order modes could be excited.

The intensity autocorrelation trace of the fs transform-limited laser pulse before and after delivery through 3 m of DC-ARF demonstrate a greater pulse stretching for shorter wavelengths **(Fig. 2e, f)**. This could be related to intermodal dispersion rather than non-linear processes taking place inside the fiber, as spectral broadening is negligible **(Fig. 2b)**. An order of magnitude higher stretching is observed in the SCF, for example at 810 nm the pulse-width was measured to be 116 fs before the fiber and 2.7 ps after the fiber.

The pulse durations calculated ($\Delta\tau_p = \frac{\Delta\tau_a}{1.41}$ for a Gaussian pulse) at different wavelengths using the autocorrelation data measured at the output of each of the fibers are presented in **Table 1**. Based on the autocorrelation results, group velocity dispersion (GVD) and its parameter D were also calculated as described in [35 and 46]. Dispersion was calculated before and after the fiber so that the effect of optical components is avoided. The results for SCF, SC-ARF, and DC-ARF are presented in **Table 2**.

*Table 1.* Pulse duration of 116 fs laser measured directly at the fiber output for SCF, SC-ARF, and DC- ARF.

|  | $\Delta\tau_p$ (fs) | | | |
| --- | --- | --- | --- | --- |
|  | **810 nm** | **850 nm** | **900 nm** | **950 nm** |
| **SCF** | 2659 | 2638.2 | 2553.2 | 2035.4 |
| **SC-ARF** | 168 | 156.3 | 133.2 | 117.2 |
| **DC-ARF** | 180 | 144 | 128 | 119 |

*Table 2.* Dispersion in fibers. Dispersion parameter (D) for group velocity dispersion (GVD) calculated for SCF, SC-ARF, and DC-ARF are shown. The error shown is the associated standard error mean (SEM).

|  | D (ps *nm$^{-1}$ * km$^{-1}$) | | | |
|---|---|---|---|---|
|  | **810 nm** | **850 nm** | **900 nm** | **950 nm** |
| **SCF** | - 103 (± 0.29) | - 90.7 (± 0.23) | -78.3 (± 0.31) | -56 (± 0.26) |
| **SC-ARF** | 0.163 (± 0.56) | 0.404 (± 0.45) | 1.45 (± 0.48) | 2.13 (± 0.39) |
| **DC-ARF** | -0.48 (± 0.29) | 1.1 (± 0.44) | 1.81 (± 0.41) | 1.8 (± 0.35) |

From the autocorrelation results, it can be concluded that the effect of dispersion is negligible over these lengths of SC-ARF and DC-ARF. Since MPM signals are inversely proportional to the pulse width, a very low level of dispersion is ideal for multiphoton endoscopy using optical fibers.

## Comparison of SCF, SC-ARF, and DC-ARF in non-de-scanned (NDD) and de-scanned (DD) detection scheme with second harmonic generation imaging

We tested all three fibers under two experimental episcopic (back-scattered collection geometry) imaging configurations used in MPM as shown in **Fig. 3** and **Fig. 4**: the non-descanned detection (NDD) tests the performance of the fiber for delivering fs laser pulses while the de-scanned detection (DD) tests both the delivery of pulses and collection of the back-scattered signal. While both could potentially work, the latter is more suitable for an MPM micro-endoscope as a bulky detector doesn't need to be part of the probe (at the distal end). We used the MPM technique of Second Harmonic Generation (SHG) imaging for demonstrating the performance of the fibers under the above-described configurations. SHG is a non-linear process sensitive to non-centrosymmetric materials wherein two photons interact to generate a new photon at twice of frequency of the incident photons. In biology and medicine, SHG imaging is predominantly used in analysis of collagen fibrillar structure and composition in many medical conditions including cancer and fibrotic diseases [47 – 49]. Collagen is the most abundant protein found in the human body [50], as such the capability to selectively image it is greatly beneficial for biomedical analysis and clinical diagnosis.

Schematics of the experimental setups are presented in **Fig. 3 and 4**. In NDD **(Fig. 3 A)**, a tunable Ti:Sapphire oscillator (116 fs, 90 MHz Spectra-Physics, Mai Tai) was used to generate pulses between 710 nm and 990 nm. After passing through a Faraday isolator the laser was coupled into each of the respective SCF, SC-ARF and DC-ARF fibers. The light was delivered to the home-built multiphoton microscope. Laser beam scanning was performed with galvanometric scanning mirrors (Cambridge instruments). A Leica DMRB upright microscope frame was used. Optical components to direct the light to the sample included a dichroic mirror (Semrock, FF458-Di02; 458 nm long pass) located before the microscope objective (Nikon 20x, 0.75 NA), which focused the laser on the sample. The backward SHG signal was collected by the same objective and reflected by the same dichroic mirror and then filtered further using a band pass filter at 405 ±5 nm (Thorlabs, FBH405-10) for 810 nm excitation and at 450 ± 5 nm (Thorlabs, FBH450-10) for 900 nm excitation. The SHG signal was detected using a photomultiplier tube (PMT) (Hamamatsu, H10722-01). ScanImage 2016b software operated by MatLab was used to control image acquisition by synchronization [51]. Galvo scanning and data acquisition from the PMT's were interfaced with the computer via a DAQ card (PCI 6110, National Instruments). We carried out experiments with both 1 m and 3 m length of fibers. 3 m length is more relevant for endoscopy and the results are shown in **Fig. 3** and **Fig. 4**. Expectedly the imaging results with the free-space coupled laser (**Fig. S1**) and through 1 m of fiber (**Fig. S2** and **Fig. S3**) have higher image quality and are presented in supplementary information.

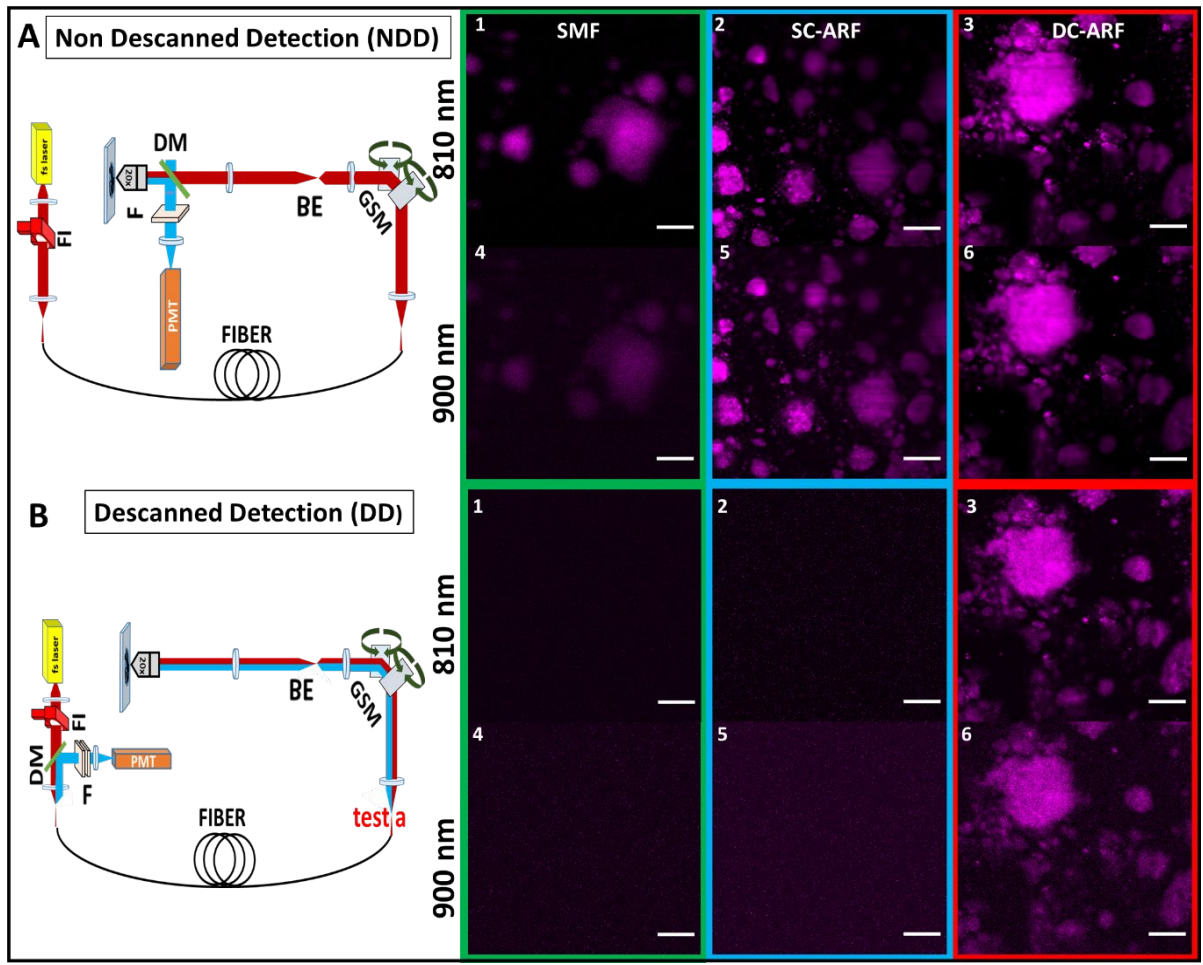

**Fig. 3 Optical setup and images of SHG-active nanoscrystals with all fibers.** Schematic diagram of experimental setups for SHG microscopy in the (**A**) NDD and (**B**) DD configuration and corresponding images acquired on a barium titanate nanocrystal sample. The excitation path is represented by red thick line, while the collection signal track is showed in blue. FI – Faraday isolator, GSM – Galvanometric scanning mirrors, BE – Beam expander, DM – Dichroic mirror, F – Filter, PMT – Photon multiplier

tube. **(1-6)** The images of barium titanate nanocrystal sample taken using NDD and DD configuration with a 3 meter long SCF, SC-ARF and DC-ARF at 810 nm and 900 nm excitation wavelength are shown (NDD: 20x objective, zoom 3, 341 pixels x 341 pixels, 10.7 µs dwell time, ~20 mW power; DD: 20x objective, optical zoom x3, 341 pixels x 341 pixels, 27 µs dwell time, ~20 mW power, average 10 frames). Scale bar = 50 µm.

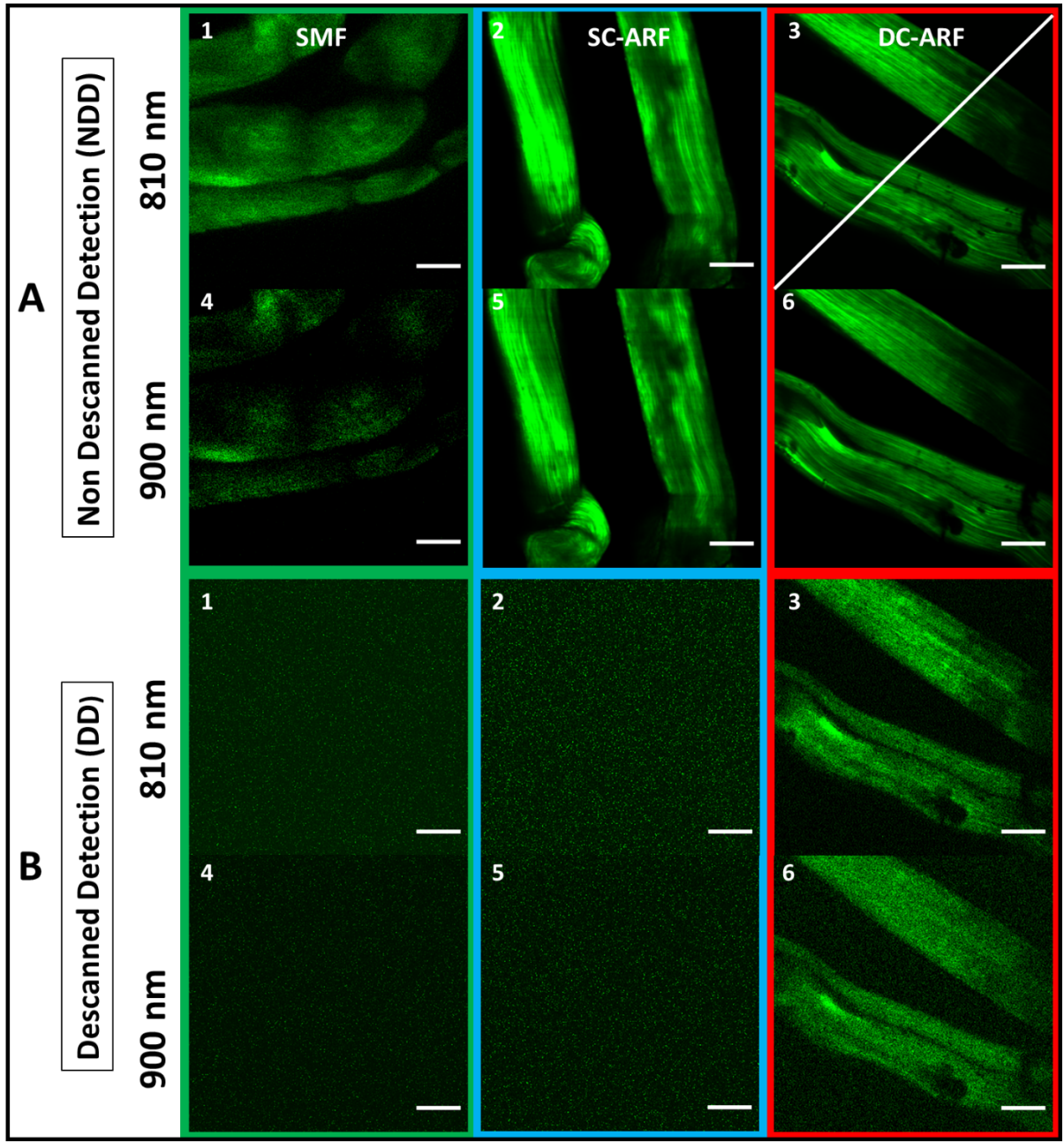

**Fig. 4 SHG imaging of biological samples with all fibers.** The corresponding images of mouse tail tendon taken using NDD **A (1-6)** and DD **B (1-6)** configuration for a 3 meter long SCF, SC-ARF and DC-ARF at 810 nm and 900 nm excitation wavelength are shown; (NDD: 20x objective, zoom 3, 341 pixels, 10,7 µs dwell time, ~20 mW power; DD: 20x objective, zoom 3, 341 pixels, 27 µs dwell time, ~20 mW power, average 10 frames). Scale bar = 50 µm.

In the case of the DD **(Fig. 3 B)**, configuration, the same setup as described above was used except that the backward collected signal was detected at the proximal end of the fiber. Hence, the dichroic mirror

(Semrock, FF548-Di02) and the detector were placed before the galvanometric mirrors. The collected signal was back coupled into the fiber core and cladding with the same lens which was used in excitation path. We measured SHG signal before the fiber but after the galvanometric mirrors (**Fig. 3B**, position 'test a' on the schematic representation of the DD setup), it was approximately 10 times weaker from that detected in non-descanned configuration. This ~10x reduced signal was coupled into to the fiber and transmitted through 1 m and 3 m long fibers for all the respective fibers, SCF, SC-ARF and DC-ARF. This weaker signal due to increased losses in the optical system and increased distance from the sample had a bearing on the image acquisition conditions and their quality (signal-to-noise ratio, SNR).

For both configurations, images were taken with an average 20 mW power on the sample. NDD images (**Fig. 3 A (1-6), 4 A (1-6) and S2 A (1-6), S3 A (1-6)**) dwell time was 10.7 μs for 1 frame while an average 10 frames of DD images were obtained with 27 μs dwell time. As the fiber were changed for comparison and hence there were slight differences in coupling into fibers that caused small changes in the field of view such that we were unable to capture images of the exact same location but largely the sample area was the same. However, power levels were measured after the objective and kept consistent between fibers. All images obtained in the NDD were compared to the images (**Fig. S1**) taken on the multi-photon microscopy system with free-space laser propagation to the same samples. Images captured via the HCFs maintained very high quality with good visualization of structural features of the samples. In the DD configuration, the capability of excitation delivery and signal collection for all fibers was investigated. Although the SCF and the SC-ARF could accommodate back-scattered signals between 400 nm 500 nm in their core region, their NA and surface area were likely insufficient to collect enough signal from the sample. Hence, no sample features were seen under our acquisition conditions and images containing only background signal were recorded for those two waveguides (**Fig. 3B (1-6), 4B (1-6), and S2B (1-6), S3B (1-6)**). However, the ability to transmit the light in two different wavelength regimes for excitation and collection, respectively, was demonstrated and proven for DC-ARF. Its large NA and large surface area of the double clad allowed it to act as a two-way waveguide. Obtained images of barium titanate crystal and mouse tail tendon indicate the versatility of DC-ARF in imaging of different sample types in an endoscopy-like scheme.

To explore this further, we performed a quantitative analysis of the signal-noise-ratio (SNR) in Fiji software (**Fig. 5, 6 and S4**) [52]. An average of grey values on multiple non-sample or blank areas, each non-overlapping area of ~6000 μm$^2$, was taken for each image. The intensity grey levels of the signal were divided by this mean value of the noise to allow the calculation of the value of SNR obtained along a line-profile drawn across images (an example is shown in **Fig. 4 A3 and Fig S4**). The graphs (**Fig. 5, and S5**) demonstrate an increase in SNR for 3 m fibers compared to 1 m fibers. The SNR in images obtained with 3 m ARFs are almost comparable to the free space laser. This may suggest that using longer waveguides for laser delivery gives better noise suppression. On the other hand, images taken with the SCF show the adverse effect of non-linearity and dispersion on the image quality. The longer length of 3 m performs worse than the shorter length of 1 m for an SCF fiber. This can be seen in **Fig. S6** that there is an almost 7-times drop in signal level between 1 m and 3 m in the SCF. On the other hand in the SC-ARF and the DC-ARF the signal levels are similar between the 1 m and 3 m lengths as can be seen in **Fig. S7**.

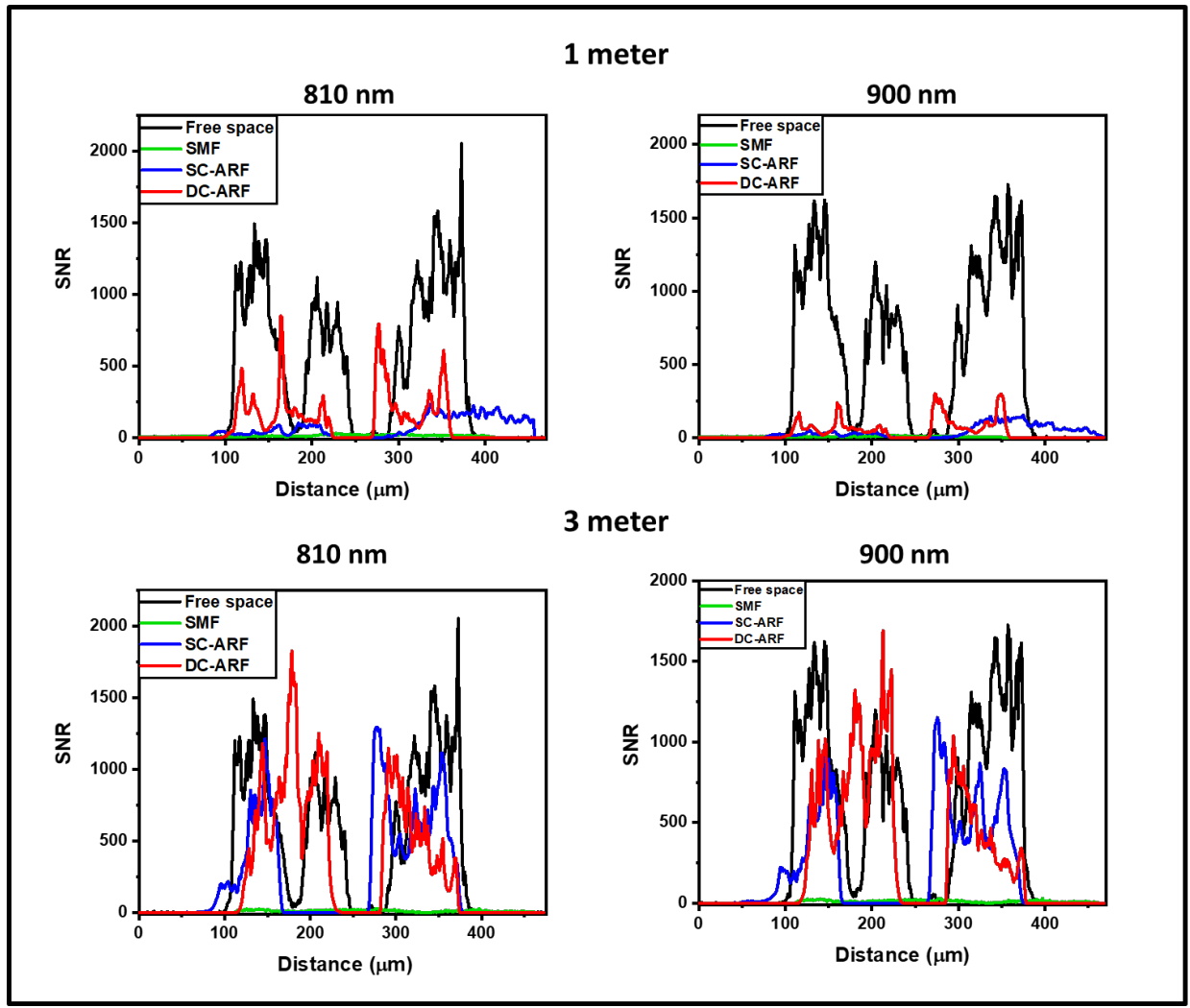

**Fig. 5 SNR analysis of images with all fibers in the non-descanned detection configuration.** Signal-to-noise ratio (SNR) profiles on images acquired in the non-descanned detection (NDD) configuration. Average SNR vs distance plots obtained for 1 and 3 m SCF, SC-ARF, and DC-ARF on SHG images acquired on a mouse tail tendon tissue in the NDD are shown at different excitation wavelengths. The SNR profiles on SHG images acquired with free-space laser coupled into the microscope are also shown for comparison. Horizontal axis represents the measured distance in μm along the line profile drawn across the image **(Fig 4 A3).**

Images in the DD configuration were only possible with the DC-ARF under our acquisition conditions as notable in **Fig. 4B**. **Fig. 6** shows SNR profile as a function of distance for all four images captured in DD scheme using 1 m and 3 m DC-ARF. All measured profiles present a sufficient SNR to see the structural details of the samples in the DD images. As in NDD, the noise suppression for 3 m DC-ARF is noticeable. This results in a similar SNR for the two fiber lengths **(Fig.5 (c, d))** as in the DD configuration the signal strength is more sensitive to distance to the detector and hence higher signal intensity is observed with the 1 m fiber.

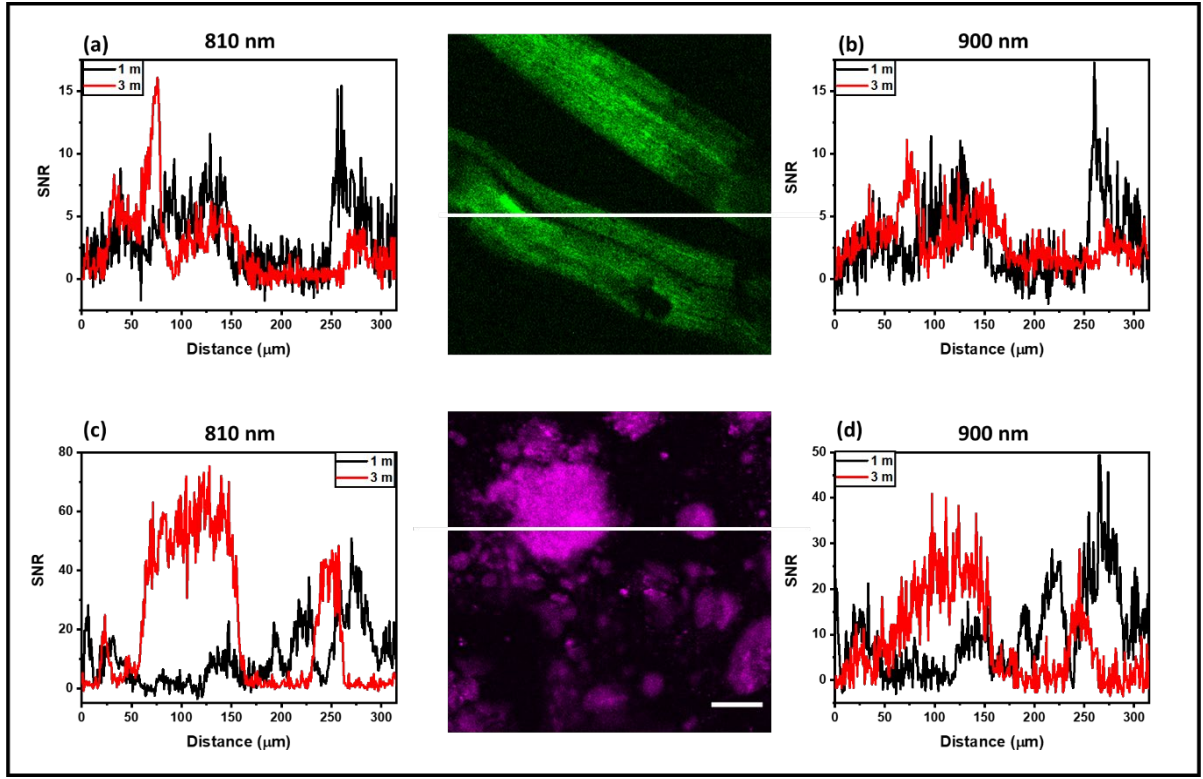

**Fig. 6 SNR analysis of images with DC-ARF in the descanned detection configuration.** Profile of the signal-to-noise (SNR) ratio in SHG images acquired in descanned detection (DD) with the DC-ARF. Images were acquired with two different length of fibers of 1 m and 3 m length. The SNR ratio from SHG images of a mouse tail tendon tissue **(a, b)** and barium crystals **(c, d)** are shown. Scale bar for the image = 50 μm. All measured profiles present a sufficient SNR to see the structural details of the samples in the DD images. Similar to the results for NDD configuration **(Fig.5 (c, d))** the noise suppression for 3 m DC-ARF is noticeable, resulting in a similar SNR for the two fiber lengths as signals are higher with the 1 m fiber.

## Conclusion

We have demonstrated a new DC-ARF as a candidate for non-linear endoscopy, able to send high power pulses and collect the non-linear generated signal from the sample. We compared the characteristics and the suitability for MPM imaging of the new DC-ARF with an SC-ARF and a typical SCF. Fabricated without any special geometry demands, such as a single coating layer that facilitates fabrication and replicability, the DC-ARF delivered ultra-short laser pulses with very low attenuation (~2 dBm$^{-1}$) and negligible dispersion and non-linearity. This was further studied in the non-descanned microscopy configuration and resulted in very high SNR images with only 20 mW power on the sample, which is advantageous from both the sample damage and imaging perspective. However, the real endoscopic test for DC-ARF was the descanned detection (DD) scheme that situates the detector at the proximal end of the fiber allowing for the most flexibility and portability of the micro-endoscope. In the descanned detection configuration, the DC-ARF's clad was utilized for backward multi-photon signal collection. In this descanned detection "endoscopy-like" scheme, the DC-ARF outperformed the SC-ARF and SCF and demonstrated a good level of signal collection via its high NA cladding. Here also a power of 20 mW incident on the sample was enough to provide high quality (>10 SNR) images. We report a systematic quantitative comparison between the new DC-ARF, a comparable SC-ARF and a typical SCF in terms of their characteristics and imaging performance. Moreover, we compare the both the descanned and non-descanned configurations with a new DC-ARF fiber, wherein the detectable signal power itself is ~10 times weaker in the former compared to latter, reinforcing the necessity to have the double-clad to improve collection. We find that a higher SNR is obtained with a longer length of 3 m of fiber compared to 1 m highlighting the benefit of using DC-ARFs for endoscopic applications. Thus, replicable, reliable and robust DC-ARF fibers such as that presented here are key for non-linear imaging based micro-endoscopic applications. Future studies will aim to demonstrate a complete micro-endoscopy system with optical fiber probe for testing clinical feasibility.

**Acknowledgements**

MS was co-funded by a studentship from the Institute for Life Sciences and ERC Grants to SM (NanoChemBioVision 638258) and FP (LightPipe 682724). Funding support from the AirGuide Collaboration fund and EPSRC grants (EP/P030181/1, EP/T020997/1) is also acknowledged. NVW gratefully acknowledges support from the Royal Society through a University Research Fellowship.

**Conflict of interest**

The authors declare no conflict of interest for this article.

*Supplementary information*

# Double Clad Antiresonant Hollow Core Fiber and Its Comparison with other Fibres for Multiphoton Micro-Endoscopy

Marzanna Szwaj[1,2,3], Ian A Davidson[1], Peter Johnson[2,3], Gregory Jasion[1], Yongmin Jung[1], Seyed Reza Sandoghchi[1], Krzysztof Herdzik[1,2,3], Konstantinos Bourdakos[2,3], Natalie V. Wheeler[1], Hans Christian Mulvad[1], David J Richardson[1], Francesco Poletti[1]*, Sumeet Mahajan[2,3]*

[1]*Optoelectronics Research Centre, University of Southampton, SO17 1BJ, UK*
[2]*Institute for Life Sciences, University of Southampton, SO17 1BJ, UK*
[3]*School of Chemistry, University of Southampton, SO17 1BJ, UK*
*frap@orc.soton.ac.uk*
*S.Mahajan@soton.ac.uk*

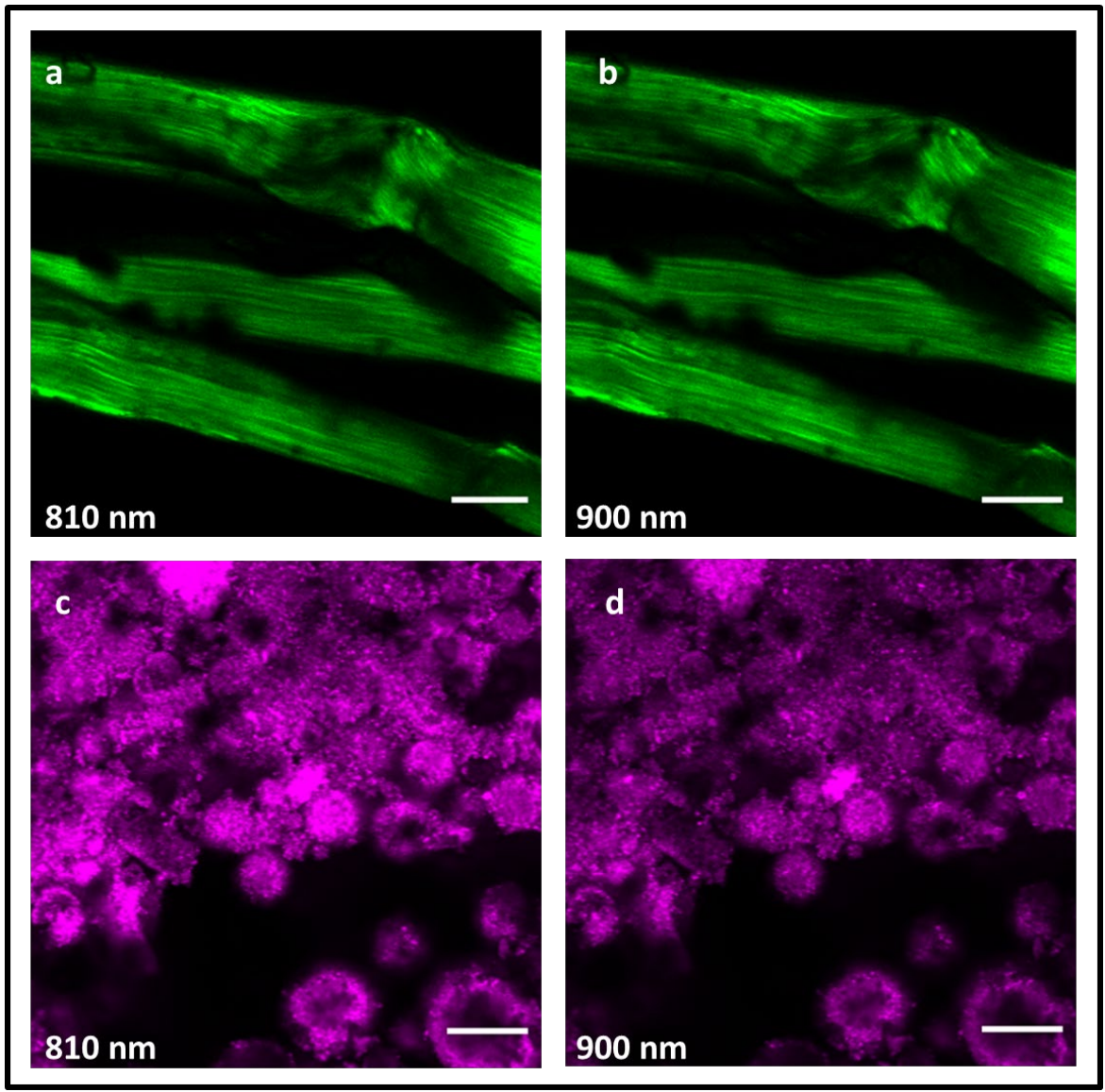

**Fig. S1** The images of mouse tail tendon (a, b) and barium titanate crystal (c, d) captured using a free-space laser propagation at 810 nm and 900 nm excitation wavelength (20x objective, zoom 3, 341 pixels x 341 pixels, 10.7 µs dwell time, ~ 20 mW power). Scale bar = 50 µm.

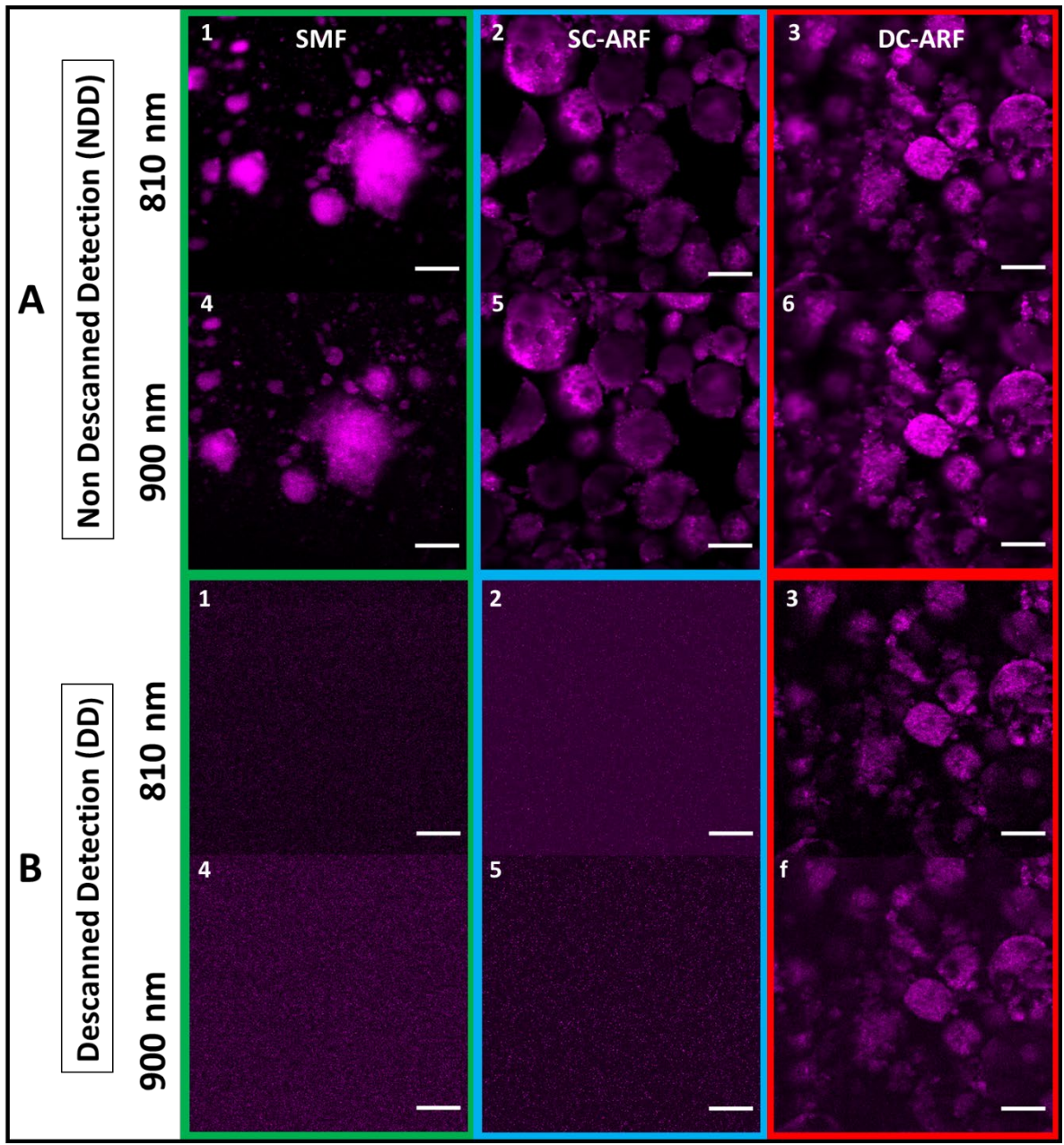

**Fig. S2** The images of barium titanate crystal taken using NDD **A (1-6)** and DD **B (1-6)** configuration 1 meter long SCF, SC-ARF and DC-ARF at 810 nm and 900 nm excitation wavelength (NDD: 20x objective, zoom 3, 341 pixels x 341 pixels, 10.7 μs dwell time, ~20 mW power; DD: 20x objective, zoom 3, 341 pixels, 27 μs dwell time, ~ 20 mW power, average 10 frames). Scale bar = 50 μm.

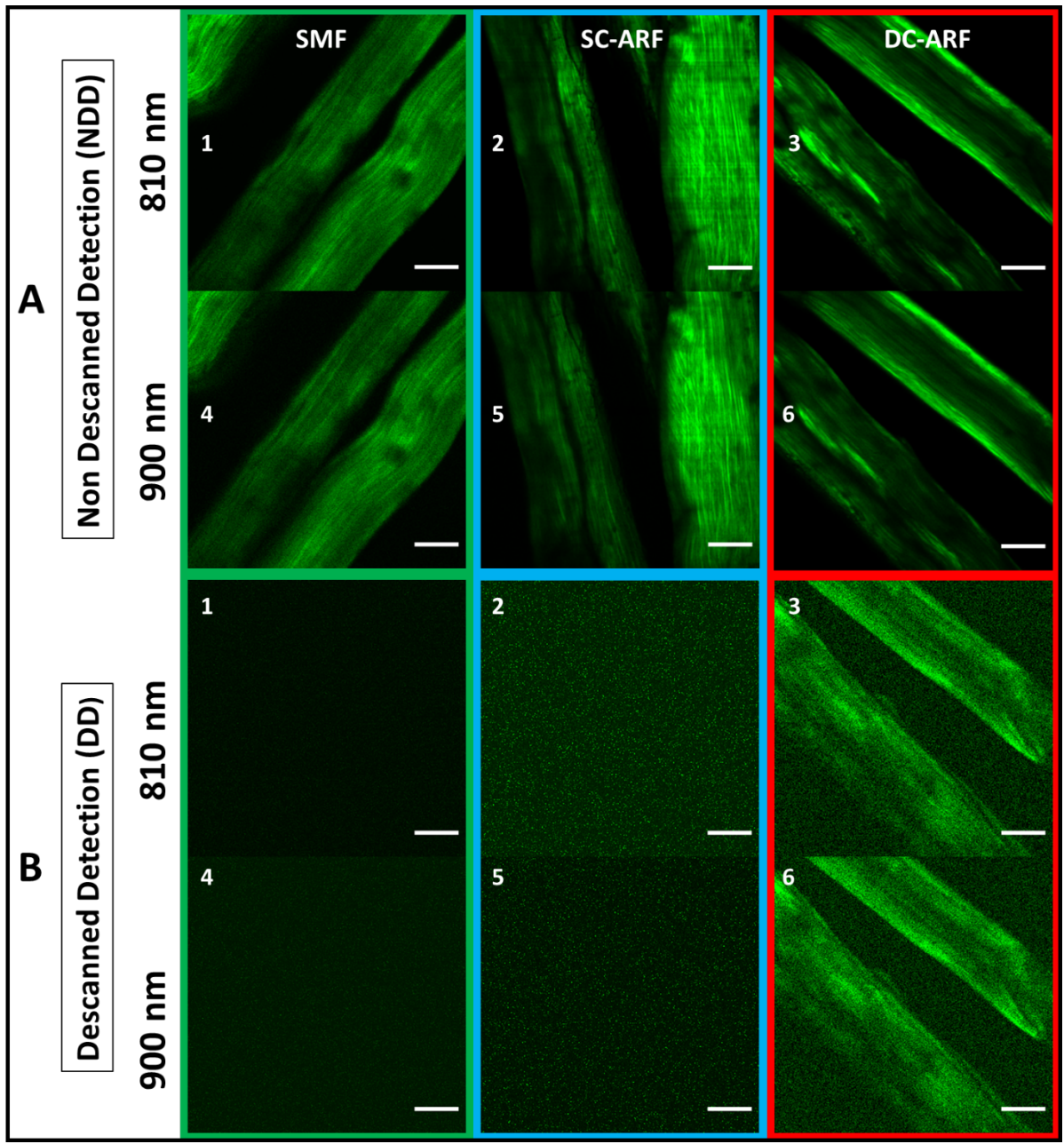

**Fig. S3** The images of mouse tail tendon taken using NDD **A (1-6)** and DD **B (1-6)** configuration (with their schematic representation) for 1 meter long SCF, SC-ARF and DC-ARF at 810 nm and 900 nm excitation wavelength (NDD: 20x objective, zoom 3, 341 pixels x 341 pixels, 10.7 µs dwell time, ~20 mW power; DD: 20x objective, zoom 3, 341 pixels x 341 pixels, 27 µs dwell time, ~20 mW power, average 10 frames). Scale bar = 50 µm.

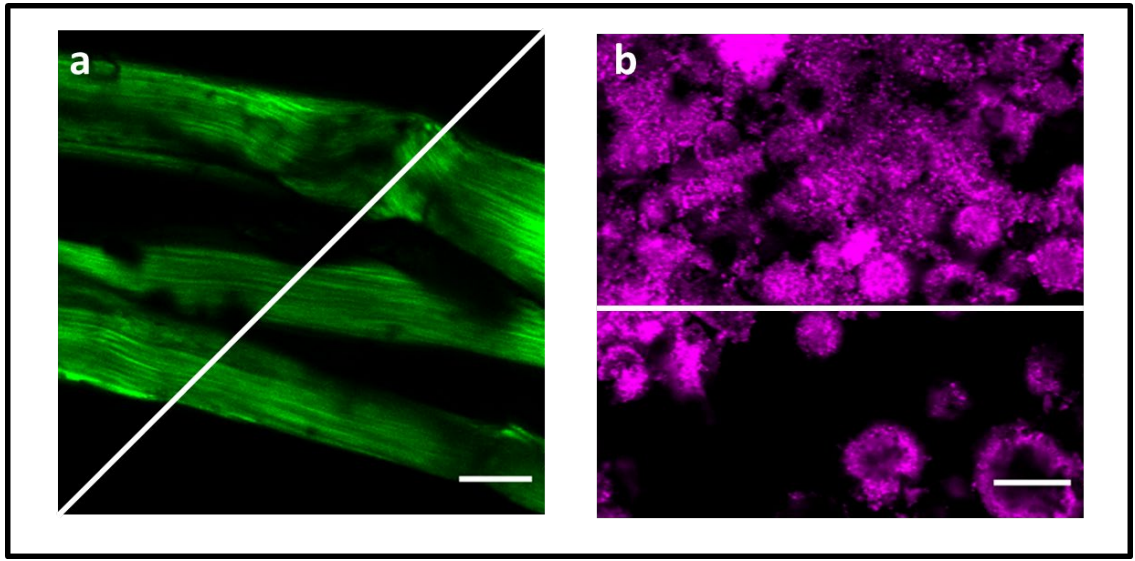

**Fig. S4** Example of mouse tail **(a)** and barium titanate crystal **(b)** images with the line profiles used in **Fig. 5** and **S4** respectively.

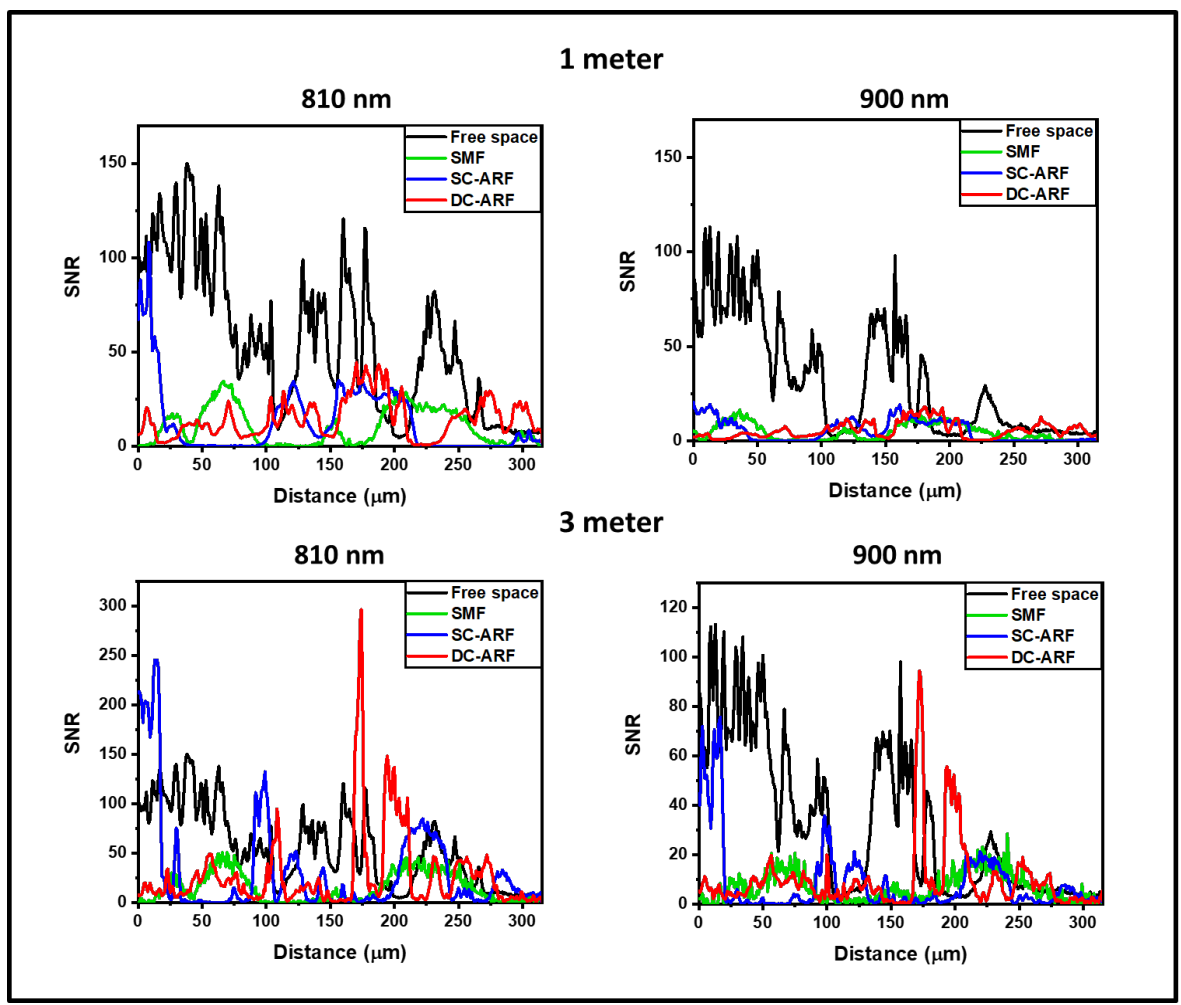

**Fig. S5** Average SNR vs distances plots obtained for 1 and 3 m SCF, SC-ARF, and DC-ARF in NDD compared with the imaging configuration with a free-space laser coupled. These SNR profiles are calculated for images of a barium titanate nanocrystal sample.

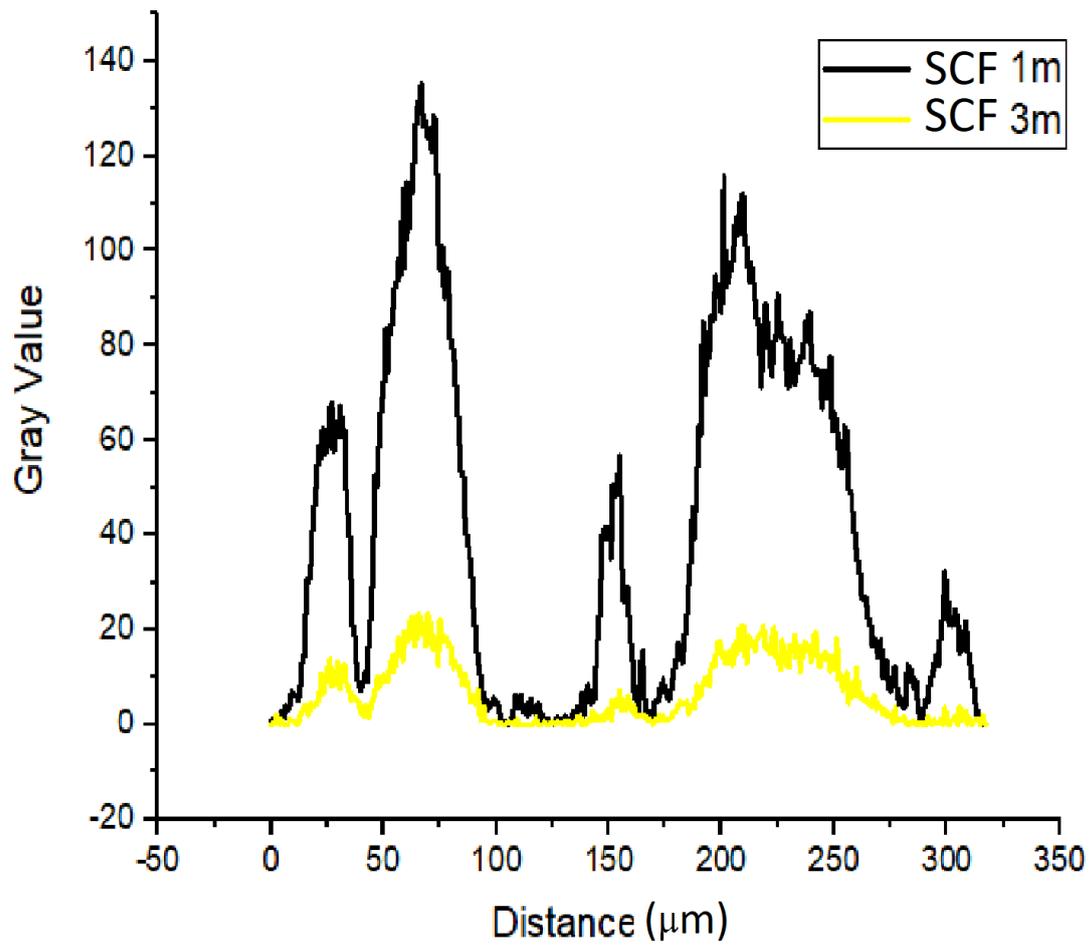

**Fig. S6** Signal intensities over a line profile of the same image for a 1 and a 3 m SCF. The signals decrease approximately 7 times due to anomalous GVD and non-linearity. The images were collected in the NDD configuration.

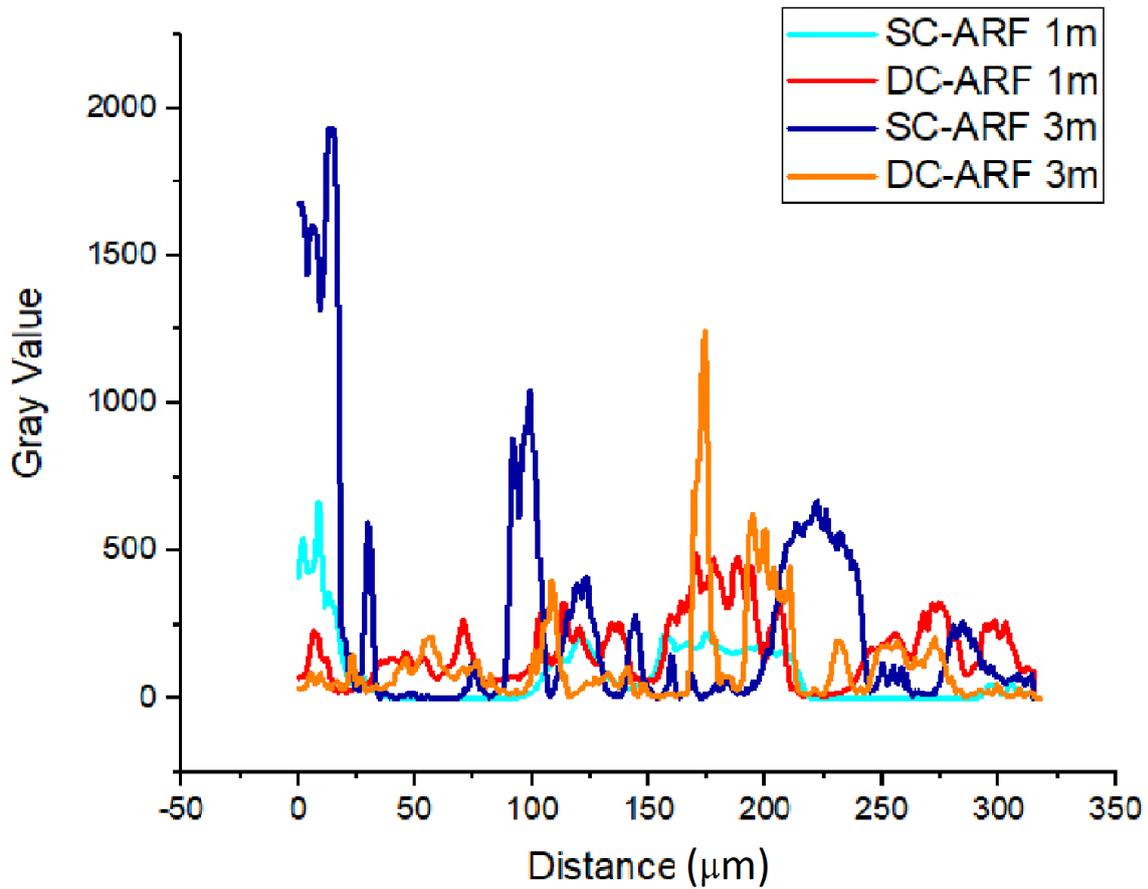

**Fig. S7** Signal intensities over a line profile of approximately the same imaging area (due to alignment issues the image was not exactly the same) for a 1 and a 3 m SC-ARF and DC-ARF. The signals are of approximately similar levels between the two lengths in either of the fibers. The images were collected in the NDD configuration.